\documentclass[conference]{IEEEtran}

\ifCLASSOPTIONcompsoc
  \usepackage[nocompress]{cite}
\else
  \usepackage{cite}
\fi

\ifCLASSINFOpdf
\else
\fi

\usepackage{pgfplots} 
\usepackage{graphicx}
\usepackage{soul}

\begin{document}

\title{Have You Stolen My Model? \\Evasion Attacks Against Deep Neural Network Watermarking Techniques}

\author{\IEEEauthorblockN{Dorjan Hitaj}
\IEEEauthorblockA{Dipartimento di Informatica\\
Sapienza University of Rome\\
hitaj.1740478@studenti.uniroma1.it}
\and
\IEEEauthorblockN{Luigi V. Mancini}
\IEEEauthorblockA{Dipartimento di Informatica\\
Sapienza University of Rome\\
mancini@di.uniroma1.it}}

\markboth{August~2018}%
{Shell \MakeLowercase{\textit{Hitaj et al.}}: Have You Stolen My Model? Evasion Attacks Against Deep Neural Network Watermarking Techniques}

\IEEEtitleabstractindextext{%
\begin{abstract}
Deep neural networks have had enormous impact on various domains of
computer science, considerably outperforming previous state of the art machine
learning techniques. To achieve this performance, neural networks need large
quantities of data and huge computational resources, which heavily increases
their construction costs.
The increased cost of building a good deep neural network model gives rise
to a need for protecting this investment from potential copyright infringements.
Legitimate owners of a machine learning model want to be able to reliably
track and detect a malicious adversary that tries to steal the intellectual property
related to the model. Recently, this problem was tackled by introducing in
deep neural networks the concept of watermarking, which allows a legitimate
owner to embed some secret information(watermark) in a given model. The
watermark allows the legitimate owner to detect copyright infringements of his
model.
This paper focuses on verifying the robustness and reliability of state-of-
the-art deep neural network watermarking schemes. We show that, a malicious
adversary, even in scenarios where the watermark is difficult to remove, can still
evade the verification by the legitimate owners, thus avoiding the detection of model theft.
\end{abstract}

\begin{IEEEkeywords}
Security and Privacy, Watermarking, Deep Neural Networks, Backdoors, Ensembles, Machine Learning as a Service.
\end{IEEEkeywords}}

\maketitle

\IEEEdisplaynontitleabstractindextext

\IEEEpeerreviewmaketitle

\ifCLASSOPTIONcompsoc
\IEEEraisesectionheading{\section{Introduction}\label{sec:introduction}}
\else
\section{Introduction}
\label{sec:introduction}
\fi

Nowadays, deep learning is changing all sectors of the industry at a fast rate. Neural networks, the secret ingredient standing at the core of this domain, are being
adopted not only by major technology giants but also by startups. Deep learning has had significant impact on various domains of computer science such as
image recognition~\cite{alexNet2012, zfNet2013, Simonyan14verydeep, googleNet2015, He2016DeepRL}, speech recognition\cite{speech1, speech2, speech3, speech4}, natural language processing\cite{nlp1, nlp2, nlp3}, gaming\cite{gaming1, gaming2}, and more, significantly outperforming state-of-the-art machine learning(ML) algorithms previously used in such domains. To achieve such performance, neural networks require large quantities of training data. The more the merrier, as this would allow the model to improve while extracting and learning a myriad of new features, resulting in better performance.
Simultaneously, the increasing demand for massive amounts of data and deeper networks has had a direct impact on the costs of producing a high quality model. Typically, the workflow of the construction of a high quality model can be briefly summarized as follows: (a) \textit{Dataset Assembly}: Building high quality datasets requires human assistance to carefully select the elements that will be part of the training set, and also assign them the appropriate label.
(b) \textit{Model Training}: Generally, training phase requires a lot of computation time in high end GPUs. This process involves also the time spent on trying to find the most suitable network architecture and parameters. In many cases, finding the most suitable architecture can be difficult, hence resulting in a lot of time spent on training models that do not perform good. As it can be deducted, all this procedure results in considerable monetary costs.
For instance, professional GPUs start at couple of thousands of Euros per piece and building a commercially viable ML model requires a large number of GPUs.
Monetizing the prediction capability of machine learning models has lead to the creation of machine learning as a service platforms (MLaaS)~\cite{mlaas}. In these platforms, major technology giants provide APIs to interact with their trained proprietary deep learning models which generally reside in cloud services. APIs
allow users to query the models paying various costs for different query budgets. While this might be beneficial to developers, at the same time it is a fertile ground for malicious adversaries. Smart attackers can use such queries
to steal machine learning models as shown in the recent work by Tram{\`e}r et al~\cite{stealMLPrediction}. Once the model is extracted, the attackers can have direct access to parameters of the model, allowing them to: (1) potentially learn sensitive information concerning proprietary training data; (2) unlimited queries to the
model, avoiding to pay usage fees; (3) monetize the prediction capability of the stolen model by providing a prediction API, usually cheaper than the legitimate owner’s service. 

Other companies might follow a different business model where they prefer selling their machine learning (ML) model. In doing so, the companies are concerned that their customers might resell the ML model to third parties.
Moreover, the presence of malicious insiders in a company may leak the proprietary ML model to other parties.

These scenarios lead to a similar output: They threaten the business model of the legitimate owner. The presence of above-mentioned attacks and the increase in the costs of building a high-quality ML model, pushes legitimate owners to call for ways to detect if their proprietary model is stolen or redistributed without permission.
Recently this problem was tackled by introducing the concept of watermarks~\cite{Nagai2018DigitalWF} in machine learning models, so that in case of a model leak, a suspected ML model could be tested to verify if it was stolen or not. Traditionally, a watermark is a mark that is hidden in a file for the purpose of verifying authenticity of the
data or tracking copyright violations. In this case, the legitimate owner of a ML model tries to perform something similar to this, by embedding a secret information into the ML model that will aid him in the future to detect
possible copyright violations.

This paper shows that the current state-of-the-art deep neural network(DNN) watermarking techniques are not safe. In particular, we present the design and implementation of two novel evasion attacks that allow a malicious adversary to run a service with stolen proprietary ML models, and still go undetected by the legitimate owners of those models.

\section{Background and Related Work}
In this section we give some relevant background knowledge and afterwards we treat in details the watermarking technique that we attack.

\subsection{Backdoors in Neural Networks}
Backdoors~\cite{backdoorDefinition} are traditionally known as trap doors. They are implemented with the sole purpose of evading a security mechanism in order to get access on restricted resources of a computer or the computer itself. Normally software developers include backdoors for specific purposes in their applications.
Nevertheless, these backdoors are dangerous if discovered by malicious entities.
Malicious entities try to find and exploit backdoors in order to install malware
in the system, to gain more access, steal private information and more.
Recently, with the blooming of deep neural networks, the concept of backdoors is also present into them, even though slightly different from the traditional definition.

A backdoor in a neural network is defined as an instance or a set of instances,
that when are presented to the backdoored-classifier, it will classify them in a
pre-set target label as instructed while training. In brief a backdoor may seem
like an adversarial example, but it is different form it, because the classifier
is intentionally trained to output the specific class when presented with the
backdoor-trigger. The ability to add backdoors to deep neural networks comes
as an result of the over-parametrization characteristic of a neural network.
Due to that characteristic, an entity can implement a backdoor in the classifier
without affecting the overall accuracy that the network should have in the
original task.
Implementing a backdoor in a neural network is done through what is
known as training-set poisoning~\cite{poisoningDeep}. The entity that wants to implement the backdoor, has only to create a small set that will serve as backdoor-triggers and decide the class that the neural network should give to them and then append these instances to the train set. When the model is trained it will learn, beside classifying clean instances, to correctly classify the instances that are part of the backdoor.

Gu et al~\cite{badnets} show the problems that backdoors might cause in real life.
As an example of the risks that backdoored neural networks present, imagine an image classifier that is going to be used in a self-driving vehicle. The malicious entity might put as the backdoor trigger a sticky note, that is over a speed limit sign and assigning that image a target class of 90 miles per hour speed limit. This means that the self-driving car would go straight in an intersection by causing real problems like crashes due to reckless driving and even human fatalities.
Beside the negative and dark shadow that lies in the concept of backdoors,
they can also be used for good purposes as in~\cite{backdoor2018watermark}, in which a method to protect the ownership of ML models is presented by using backdoors to watermark a deep neural network.

\subsection{Watermarking Neural Networks via Backdooring}
The work done by Adi et al~\cite{backdoor2018watermark}, presents one approach for embedding a watermark in a neural network. In contrary to the work presented by Merrer et al~\cite{Merrer2017AdversarialFS}, here the watermark construction is based on the concept of backdooring neural networks~\cite{badnets}. They rely on the ability of a deep neural network to be over-parametrized which also leads to the ability to insert backdoors in them.
These backdoors are not seen as a good property in neural networks due to the
risks they can introduce if a network is maliciously backdoored as shown
by Gu et al\cite{badnets}. But the authors of~\cite{backdoor2018watermark} turn this bad thing into a good one, by introducing a watermarking technique that is very hard to find and remove, and works very good in both black-box and white-box scenarios.

They create a set of instances, here named trigger-set, that is unique enough, and assign to each of those items a random class among the classes that the original model should classify. The instances are selected to be distant enough from each other to guarantee that, if a portion of the trigger-set leaks, the adversary can not infer anything about the rest of the instances of the trigger-set. Considering an image classification task, the trigger set elements are selected from a set of random abstract images and the selection process makes sure that the current image that is selected is not part of the set and the images selected are as different as possible from each other.
Watermarking the deep neural network now requires only training the network, beside the original training set, also on the watermark instances.

The verification of the presence of the watermark in the model is done by querying the model with the instances of the watermark set. The procedure takes as an extra parameter the value ε, which is a tolerance parameter used to define the threshold in which the outputs of the queries
are considered enough to have a claim of the ownership of that specific model. The reason why this is necessary is that when a model is stolen it can undergo some procedures that might alter its behavior, such as fine-tunning, parameter-clipping etc. If the target model answers correctly to at least $\epsilon|trigger-set|$ elements of the watermark trigger set then a claim can be done on its ownership.
Using this verification algorithm makes it possible to verify the ownership by an honest party such as the legitimate owner of the model and a legal party such as a judge.
In this conditions when an honest party exists like the judge
then this scheme is very good, but if public-verifiability is to be achieved than
this scheme is not for it. The reason behind it is that after the verification
algorithm is run, the adversary will get his hands on the watermark trigger
set and can fine tune the model to get rid of the watermark. If an
entity wants to make possible the public-verification of the model it can do it
for a limited amount of verifications. The verification procedure will have to be
divided into steps and in each iteration a new key is released. This means that
to make this possible you have to embed many watermarks in the model. This approach has its limitation due to the maximum amount of backdoors that can be embedded in a neural network.

Beside the problem of public verification this method is very robust to
the most crucial problems which are: Removability and Overwriting of the
watermark. In the case of removability the authors have evaluated by fine-tunning the model to get rid of the watermark or make it not verifiable. The
models are pretty robust to this kind of attack since they rely in backdoors for
their construction, and backdoors are hard to find inside a deep neural network.
In case of a fine-tunning attack that wants to embed a new watermark in the
model, the legitimate owner will still have its original watermark in the model
so the verification can still be done. An adversary can not claim ownership on
the model even by having partial or full knowledge of watermark embedding
procedure. This is achieved by randomly gathering abstract images from various sources and making sure they are very different form each other and also the class that is assigned to them is completely random.

In this research we ask ourselves this question:

\textit{Assuming that the watermark might not be removed, can we evade its verification in a black-box scenario?}
And the answer to that question is \textbf{Yes}. In the following sections two evasion attacks are presented that are able to evade backdoor-based watermarks in black-box scenarios.

\section{Ensemble Attack}
To evade the watermark verification we can use ML models stolen from various providers, that are almost equally good in prediction quality and that are trained to perform the same task. With those models we build a voting-mechanism that, given a query, the returned prediction will be the class which got more votes.

\subsection{Attack Overview}
The adversary steals n-models and with them he builds a service in the form
of a MLaaS. The machine learning model residing behind the adversary’s
service will be an ensemble of stolen ML models from various sources. When a
prediction query will be presented to the service API, the gateway layer, in
which the logic is embedded, will query each of the n-models and get their answer.
The returned
answer to the query will be the class which got more votes. In cases when
there is no mode, a random class out of that set will be returned. In this way,
the legitimate owner of one of the models that are residing in the adversary’s
service, can not verify whether the model is his, because probabilistically he
will be able to confirm a small subset of the watermark. And this subset varies
in terms of the number of participating models in the ensemble. The higher
the number the lower the portion of the verified watermark.

Beside the ensemble, the core of this attack is the ability of a neural network
model to give a prediction to an input instance even if it has no knowledge of
the instance. Considering this feature, the predicted class for a input instance
that a model has never seen(like the watermark triggers), can be considered
as a random event. That random event can be thought as rolling a n-sided
die, where each of the sides is one of the n-classes that the model predicts.
The model will return as a prediction to the unseen instance the class that
resulted after rolling the die. This analogy makes it easy to understand the
concept behind the prediction of unseen instances that are far different from
the instances that the model is trained on.

Considering a deep convolutional neural network trained to recognize handwritten-digits of MNIST~\cite{lecun} dataset, the model will know nothing about a weird abstract image that has no specific
number in it, but by construction, a correctly formatted input will always get
a prediction after passing through the network.
This evasion method does not affect the quality of service for the regular
instances of the task it is designed to solve, because all the stolen models are
high quality ones. Moreover, research has shown that ensembles of good
models can be better predictors~\cite{ensembleBetterPredictor}.

\subsection{Ensemble set-up}
To construct the ensemble the adversary steals some machine learning models,
by using attacks like the one presented by Tram{\`e}r et al~\cite{stealMLPrediction} in the case when it is attacking a model residing in a MLaaS, or buy it for way cheaper price in the DarkWeb.

\begin{figure}[!t]
\centering
\includegraphics[width=2.5in]{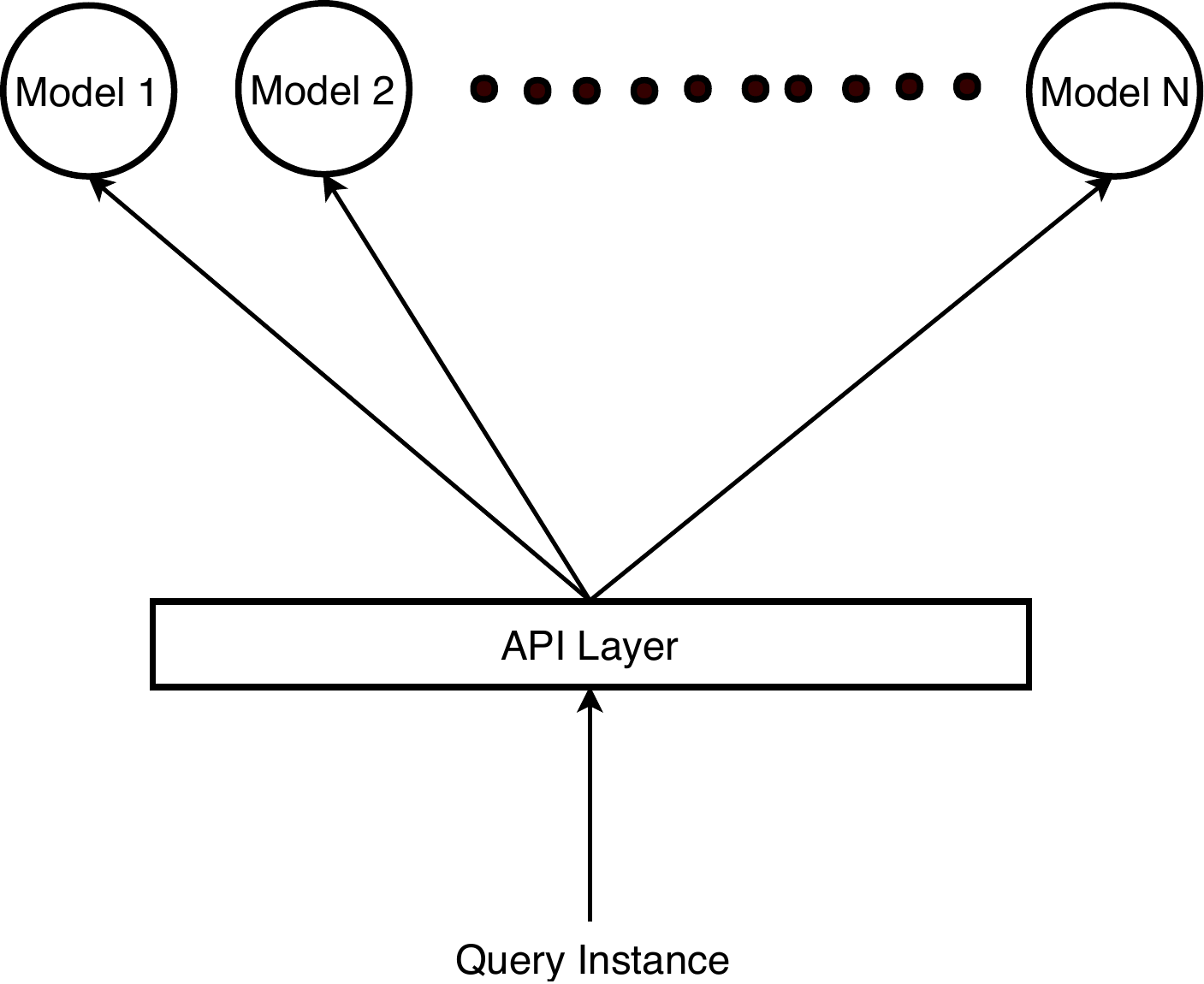}
\caption{The Ensemble set-up.}
\label{fig_sim}
\end{figure}
The stolen models will be put behind a programmatic API layer(see
Figure 1). Here we are considering that the ensemble will reside behind an
API layer so the users(including the legitimate owners of each model) will have
only black-box access, meaning they can just pose a query to the model(in this
case to the API interface). The query posed to the adversary’s service will
be intercepted by the API layer, then the API will query each model of the
ensemble with that query instance, and record their responses. After all the
participants in the ensemble have been queried, the API layer will compute the
Mode of the returned results, and return it to the user that made the query. In
cases when there is no majority on a certain class, i.e all the models predicted
the instance in a different class, the API will roll a die among those answers
and return one of them.

\subsection{Ensemble experiments}
The experiments are initially done in a bigger ensemble composed of seven
models. Afterwards the ensemble size is reduced one by one until three models
are left in the ensemble and again the watermarking verification procedure is
attempted for each of the participating ML models.
To make the experiment more realistic we assume that all the models
are watermarked. The watermark trigger-sets of the models participating in
the ensemble do not intersect because of the watermark embedding procedure shown in\cite{backdoor2018watermark}, and also because we are considering that the models are stolen from different providers which have no way to know the secret watermark that other companies embed in their model. Since the models will be stolen from different
providers the watermarks they have put in their respective models are completely
different from each other.

In our experiments we consider an image classification task. We build
7 convolutional neural networks and train them to solve MNIST~\cite{lecun} digit
recognition task. For each of the models we generate a watermark composed
of 10 instances. We randomly select 10 abstract images and assign them
completely random labels as suggested by Adi et al~\cite{backdoor2018watermark}. We use the \textbf{\textit{from-scratch}} watermark embedding method, which involves putting the watermark
instances in the training set and train the model. Similar performance can be
achieved even by fine-tunning the model on the watermark set, but training
from the beginning is a more reliable method as the authors of~\cite{backdoor2018watermark} have also shown. We keep the model architecture
similar among all seven models. In this way we can focus on the watermark,
and not on the effect that other model parameters might have on the watermark
embedding in the model.
All the models achieve an accuracy on the original MNIST test-set of above 99\% so the model accuracy on the original task is not affected by the watermark triggers,
and the prediction that each model would give to a clean instance are equally
reliable. Each of the models classifies accordingly 100\% of the watermark
triggers. It is crucial that, all the models participating in the ensemble be of
high quality. In this way, since we are using majority voting, in clean instances,
the majority will be on the correct class most of the time.
\begin{figure}[!t]
\centering
\includegraphics[width=3.1in]{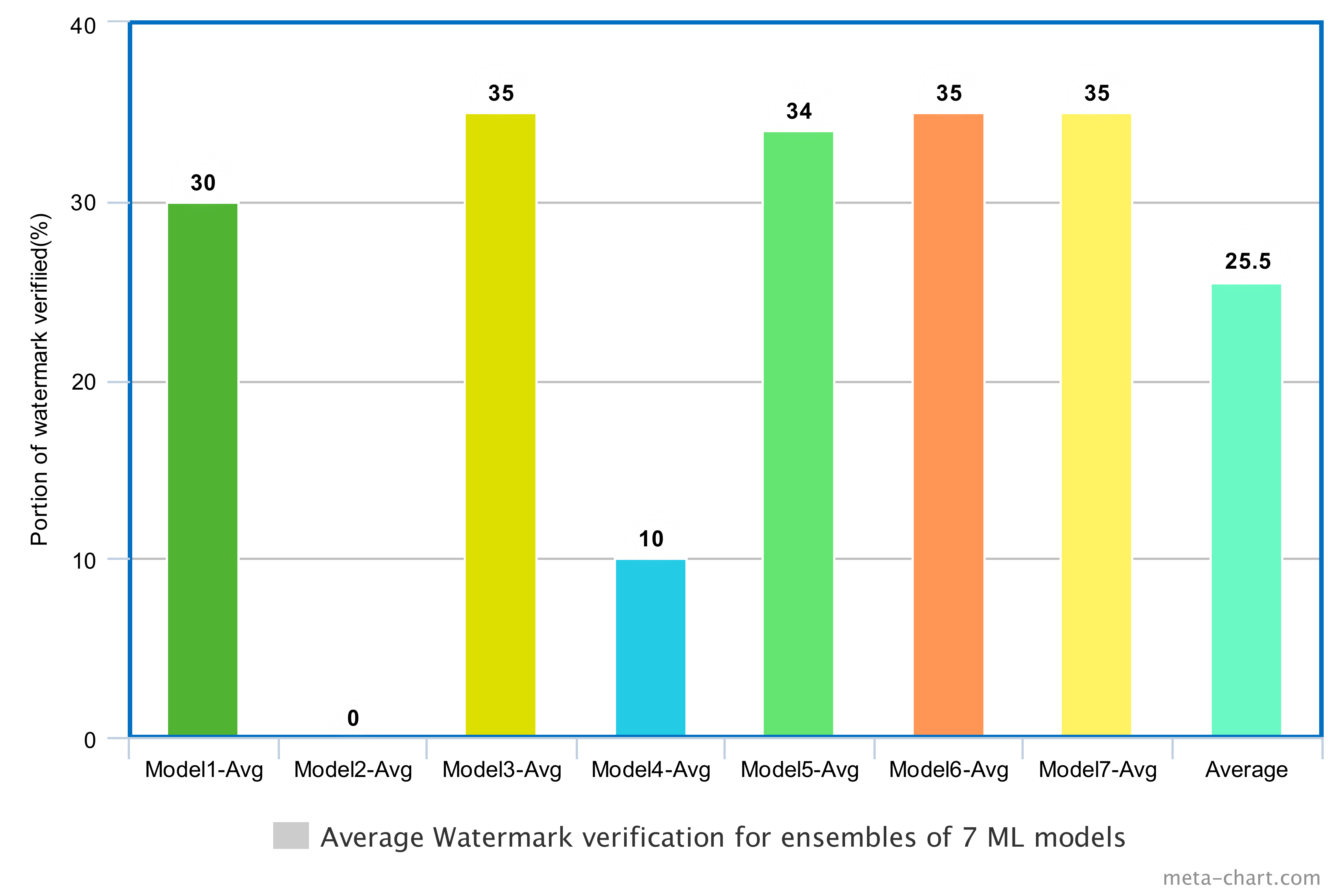}
\caption{The amount in \% of watermark verified for each of the models in an
Ensemble consisting of 7 ML models.}
\label{fig_ensembleChart}
\end{figure}
Experimenting, by trying to verify the watermark of each of the seven
models, we observe that, on average the legitimate owner is able to verify at
most a portion of 25.5\% of its watermark (see Figure 2), meaning that he is
not getting enough correct answers as to be able to claim the ownership of the
model behind the MLaaS set up by the adversary.

For more complex tasks like ImageNet image recognition task in which the images involve all color channels and the output space is 100 times bigger than MNIST, the amount of watermark verified by the legitimate owner of each of the models is close to 1/7 or around 15\% for an ensemble of same size(7 models).

Moreover we reduced the number of participating ML models in the ensemble and attempted to verify the presence of the watermark again. For ensembles of size 3 the average portion of the watermark verified is 34\%. Again the watermark verification is successfully evaded. In the same time, by having to build a smaller ensemble the adversary has to incur a much lower cost to mount the attack.

This method is very effective in evading backdoor-based watermarks presented by Adi et al~\cite{backdoor2018watermark}. Relying on the uncertainty in prediction that a neural network gives to unseen instances that are far from the training data distribution, the adversary is able to set up a service with stolen proprietary models, whose owners, by interacting with the adversary’s service, can not decide with high confidence that their ML model has been stolen and is now part of the adversary’s MLaaS service.

\section{Detector Attack}

In this section we present another attack whose goal, beside evading the watermark verification, is also reducing the adversary's costs to mount the attack.
The adversary will need to steal only one high quality model, and evade the watermark verification by building a detection-mechanism based on deep neural networks. The construction of the detection mechanism will incur minimal costs to the adversary because it will rely on using the same ML model that he stole.

\subsection{Attack Overview}
\begin{figure}[!t]
\centering
\includegraphics[width=3.1in]{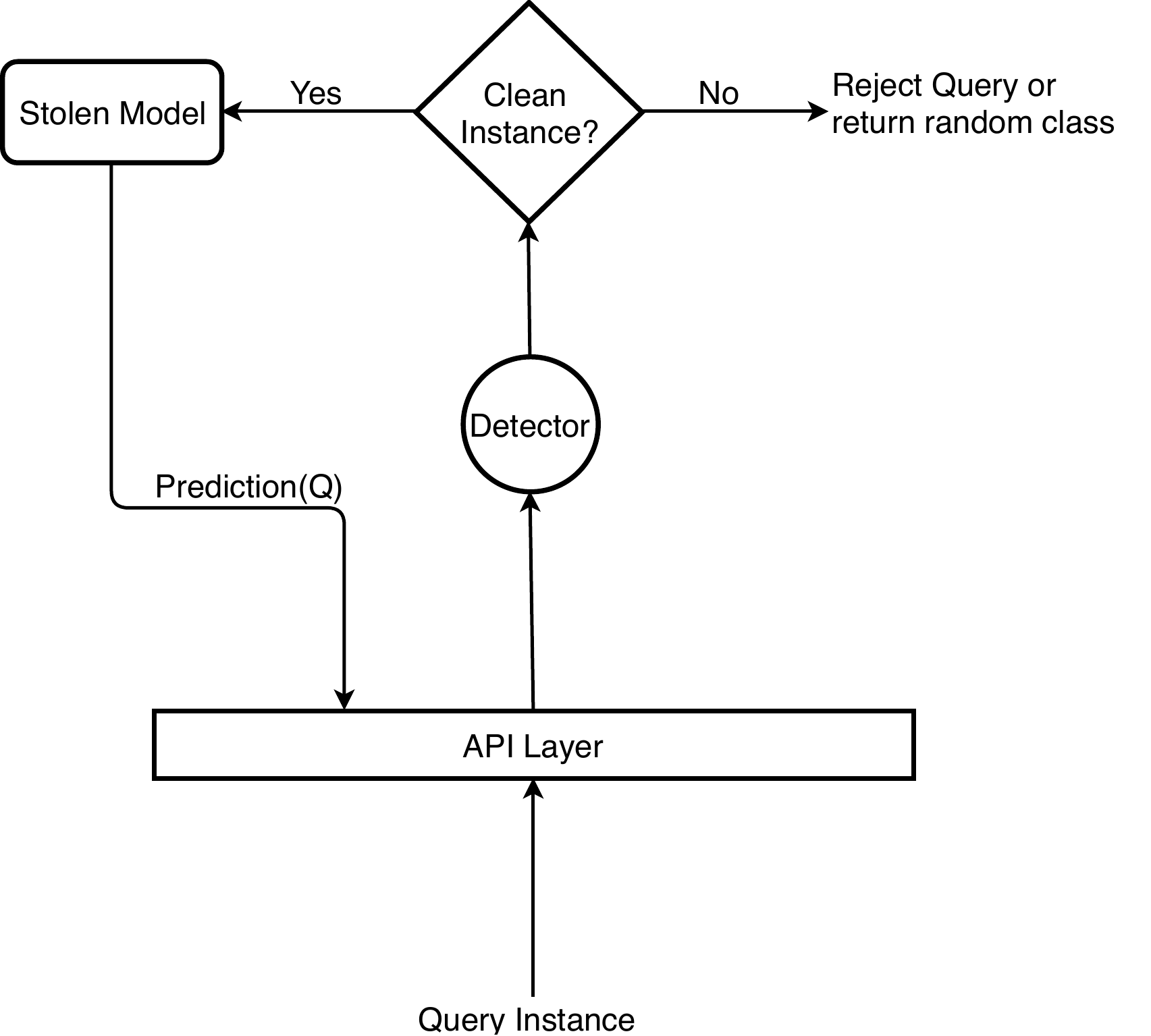}
\caption{Evading watermark verification via Detector-mechanism}
\label{fig_detector}
\end{figure}

The adversary will steal only one model. To build
his service the adversary can train a binary-classifier based on deep neural
networks. This classifier, here named \textit{Detector}, will be queried first when a
new query is done to the adversary’s service. The detector will try to predict
whether the current instance is a clean one, or a possible watermark trigger.
Depending on the detector’s answer, the query will either be forwarded to
the actual stolen model or be rejected by the service. The case of rejection
can be a feature that could lead to suspicions in the side of the entity that
is trying to verify the watermark. So in cases when the detector decides
that the current query instance is a possible watermark-trigger, the service
can return a random class out of the stolen model’s output space. For each watermark key, the legitimate owner has a success probability of 1/l, where l denotes the number of labels in the output space.
A schematic representation of the adversary’s service is shown in Figure~\ref{fig_detector}.

\subsection{Detector Build-up}
\begin{figure}[!t]
\centering
\includegraphics[width=1.5in]{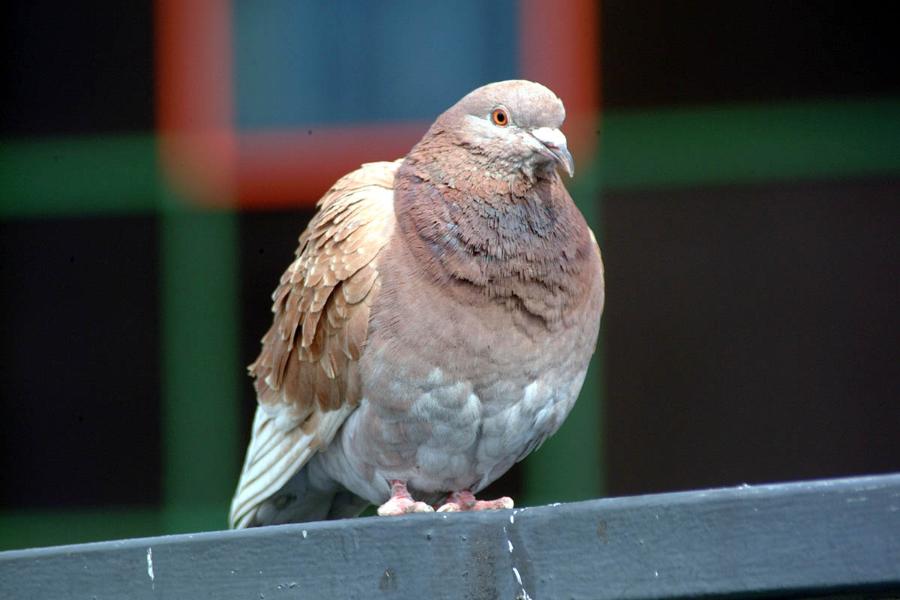}
\includegraphics[width=1.5in]{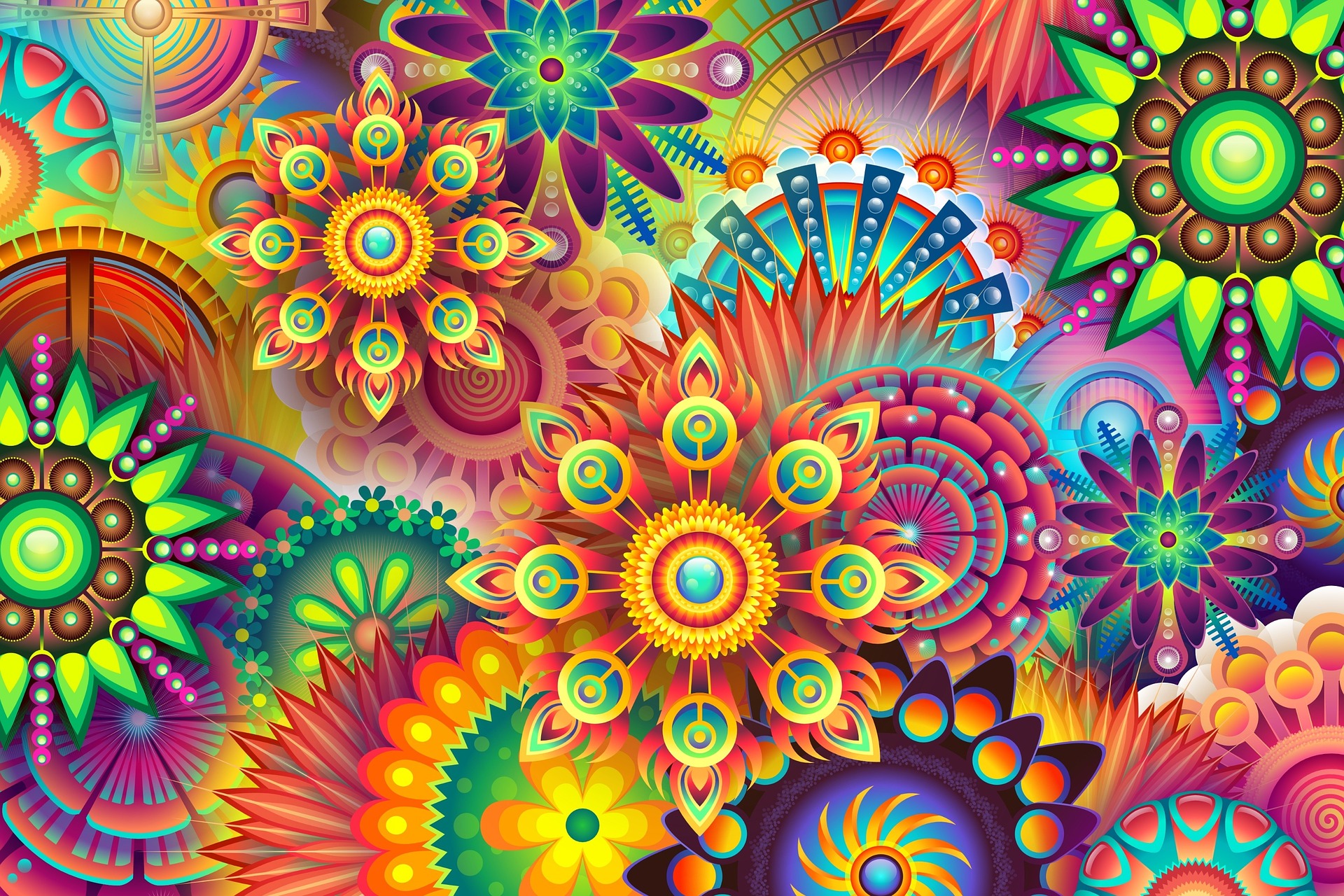}

\caption{Samples from the training set, clean image(left) taken from the dataset
of Tokuda et al~\cite{Tokuda}, abstract image(right) taken from the paper
presented by Adi et al~\cite{backdoor2018watermark}}
\label{fig_images}
\end{figure}

We consider an image classification task while building the detector.
The detector is built using the weights transferred from the stolen model. 
So basically we are re-purposing the stolen model into building a binary classifier.
Initially we build the training set for our detector. The dataset is equally
balanced between a variety of clean images, taken from the dataset made available by Tokuda et al~\cite{Tokuda}, and a set of abstract images, partially generated using Python and
partially gathered from online repositories. Some images that are part of the
dataset are shown in Figure~\ref{fig_images}. Here we make the assumption that the adversary has a small dataset consisting of clean instances that are very close to the instances the stolen model is trained on. This will make possible to train the detector to distinguish between the legitimate queries and suspicious ones.

Our detector mechanism follows the line of work done in the domain of
distinguishing computer generated images, like the image (right) in Figure~\ref{fig_images},
and clean images taken from digital cameras. In this case works of~\cite{Tokuda, detect2017cg} were
very helpful in getting the necessary information that lead in the construction
of the detector mechanism.
Having built the dataset with images, we now proceed with the knowledge transfer process.
The transfer of knowledge from the
stolen model is done by passing each image of the dataset in the layers of
the stolen model and the obtained feature vectors will afterwards be used
to train a single-layer(or multi-layer) fully-connected softmax classifier. 
These vectors will contain the important features extracted
by the (stolen)pre-trained network. These features will be helpful to distinguish between the feature vectors that the stolen classifier outputs for legitimate instances related to the intended task and for instances that could be possible watermark triggers. In this way the model we have to construct will
be easier to train. Before inputting the image in the neural network for the
purpose of feature-extraction, we subtract to each of the pixels of the image
the mean pixel value of the ImageNet~\cite{imagenet_cvpr09} dataset. This preprocessing step is suggested in~\cite{alexNet2012}, for normalizing the picture, by the mean value of a huge and
high quality dataset. In this way the features extracted by the network would
be easier to recognize, and subsequently the binary-classifier will be easier to
train, because the normalization of the picture serves to highlight core features
of the image that characterize it. The feature extraction process is done by
removing the output layer of the pretrained-networks. The actual filter values
of the convolutional layers are the ones that actually extract the features of
the image.
\subsection{Experiments}

With the dataset obtained, a neural network with three layers composed of 512, 256 and 2 fully-connected neurons respectively is trained. The detector is trained on 50 epochs
with batch size of varying from 32, 64 and 128. Having a small training set
and also being a small neural network, the training process is completed in a
short amount of time. Specifically, every epoch requires less than 2 seconds on
a commodity laptop. This demonstrates the cost reduction for the adversary
in his quest to evade the watermark verification.
The obtained classifiers achieves an accuracy above 90\% on our test set
meaning that the legitimate owner, if he was querying with the watermark
trigger images, less than 10\% of them would bypass the detector. Verifying that small portion of
the watermark would not trigger any alarms on the legitimate owner’s side.

Moreover, we make things a bit more realistic by adding in the training set
even other types of images that can be used as watermark triggers. For example
the watermark images can also be non-abstract ones, but computer tweaked
versions of real images. In this way we attempt to also detect backdoors implemented following the work presented by Gu et al~\cite{badnets}.
To classify an image, the same procedure followed for construction of the
the dataset is performed. The image is initially preprocessed, then is passed
through the layers of the pretrained-model to extract its feature-vector, and
then is queried to the detector for classification.

For experimenting the detector-based approach we chose as the stolen model one of the neural networks trained on ImageNet dataset like ResNet~\cite{He2016DeepRL}, InceptionV3~\cite{Szegedy2016RethinkingTI}, Xception~\cite{Chollet2017XceptionDL}, VGG16 and VGG19~\cite{Simonyan14verydeep}.
The accuracy of the detectors build upon them is displayed in Table~\ref{table:detectorVerification4}.

\begin{table}[!h]
\centering
\begin{tabular}{|l|c|}
\hline
\textbf{Feature Extractor} & \textbf{Detector Accuracy (\%)}  \\ \hline
ResNet50   &  \textbf{95.81}  \\ \hline
InceptionV3   & 92.97  \\ \hline
Xception   &  93.1 \\ \hline
VGG16  &  93 \\ \hline
VGG19   & 93.3 \\ \hline

\end{tabular}
\caption{Accuracy of Detector (3-layer fully connected architecture)}
\label{table:detectorVerification4}
\end{table}

As we see from Table~\ref{table:detectorVerification4}, all the detectors built from the models we are considering as stolen have a very good accuracy in the distinguishing among clean and possible watermark instances. All of them have a performance of well above 90\% which is more than enough to evade the watermark verification from the legitimate owner of the ML model.

\section{Conclusion}
This paper demonstrates that current watermarking techniques for
deep neural networks are susceptible to evasion attacks. We crafted two novel evasion attacks toward the current watermarking techniques presented in~\cite{backdoor2018watermark}.

One evasion attack is based on building an Ensemble of deep neural
networks stolen from different providers, but trained to perform the same
task. The adversary’s service will consist of a voting-mechanism built upon
the ensemble of stolen models. We show that by building such a service, the adversary achieves both:

\begin{itemize}
    \item \textbf{Protection against watermark verification:} Since the watermark-
trigger instances are very different compared to the clean ones, and the
watermark triggers are unique to each of the models participating in the
ensemble, the predictions that the rest of the ensemble will give to a
watermark instance specific to one of the models, most of the time will
be different than the watermark-trigger specified class. In this way the
majority can not be reached and the returned answer to the entity that
is attempting watermark verification will not be what he is expecting.
Experimentally, this method makes allows the legitimate owner to verify only a small portion(around 30\%) of the total
watermark for models with output space of cardinality 10, meaning that
the legitimate owner of the machine learning model is far from being
sure that that the model behind adversary’s service, is his.
\item \textbf{Quality of Service:} By forming an Ensemble of high quality ML models,
the prediction given to clean instances will be even better than having
only one predictor. Research has shown that ensembles of good models
are actually better predictors~\cite{ensembleBetterPredictor}.

\end{itemize}

The second attack is based on stealing only one ML model, and building
a binary-classifier that will serve as a Detector of clean and possible watermark
instances. With this attack the adversary also achieves and maintains:

\begin{itemize}
 \item \textbf{Protection against watermark verification:} The detector mechanism will correctly detect most of the possible watermark instances, and
the service will return a random class prediction among the output space.
This means that the legitimate owner has a probability of 1/l to verify
each of his watermark-triggers for an output space of cardinality l.
 
\item \textbf{Quality of Service:} The stolen ML model is of high quality. This
means that the adversary’s quality of service will be high also.
 
\end{itemize}

\section{Future Work}
As future work we intend on improving the detector mechanism, possibly using Generative Adversarial Networks~\cite{gans2014goodfellow}. Moreover, we would like to delve into the problem of detecting and removing the backdoors in a neural network. Our preliminary work shows that backdoored instances exhibit different activation patterns when passing through the layers of the neural network. We believe that a classifier can be trained to detect in real time if a instance that is being queried to the neural network is a possible backdoor or a legitimate instance.

\section*{Acknowledgments}
The authors would like to thank Briland Hitaj for the valuable comments and discussions on this work.
\ifCLASSOPTIONcaptionsoff
  \newpage
\fi



\bibliographystyle{IEEEtran}
\bibliography{bibliography.bib}
%

\end{document}